\documentclass[letterpaper,10pt,twocolumn]{ASPE_ExtendedAbstract}
\usepackage{setspace}
\usepackage{color}
\usepackage{cite}
\usepackage{units}
\usepackage[pdftex]{graphicx}
\usepackage{float}
\usepackage[cmex10]{amsmath}
\usepackage{balance}
\usepackage{subfig}
\usepackage{times}
\usepackage{graphicx}
\usepackage{amsmath}
\usepackage{amssymb}
\intextsep=\lineskip

\title{ Transcending the Acceleration-Bandwidth Trade-off: Lightweight Precision Stages with Active Control of Flexible Dynamics}

\author{Jingjie Wu and  Lei Zhou}
 
\affiliations{Walker Department of Mechanical Engineering\\
University of Texas at Austin, TX, USA\\
\\
}

\begin{document}
\maketitle 
\bibliographystyle{vancouverASPE}

\section{Introduction}

Micro/Nano-positioning stages are of great importance in a wide range of manufacturing machines and instruments such as wafer/reticle stage in photolithography scanners and IC inspection systems, and their motion performance is critical to the quality and throughput of the manufacturing/inspection process.
In recent years, the drastically growing demand for higher throughput and reduced power consumption in various  IC manufacturing equipment calls for the development of next-generation precision positioning system with unprecedented acceleration capability while maintaining exceptional positioning accuracy and high control bandwidth \cite{oomen2013connecting}. Reducing the stage's weight is an effective approach to achieve this goal. 
However, the reduction of stages' weight tends to decrease its structural resonance frequency, which limits the closed-loop control bandwidth and can even cause stability issues. 

\begin{figure}[t!]
\centering
\subfloat{
\includegraphics[trim={0mm 0mm 0mm 0mm},clip,width =1\columnwidth, keepaspectratio=true]{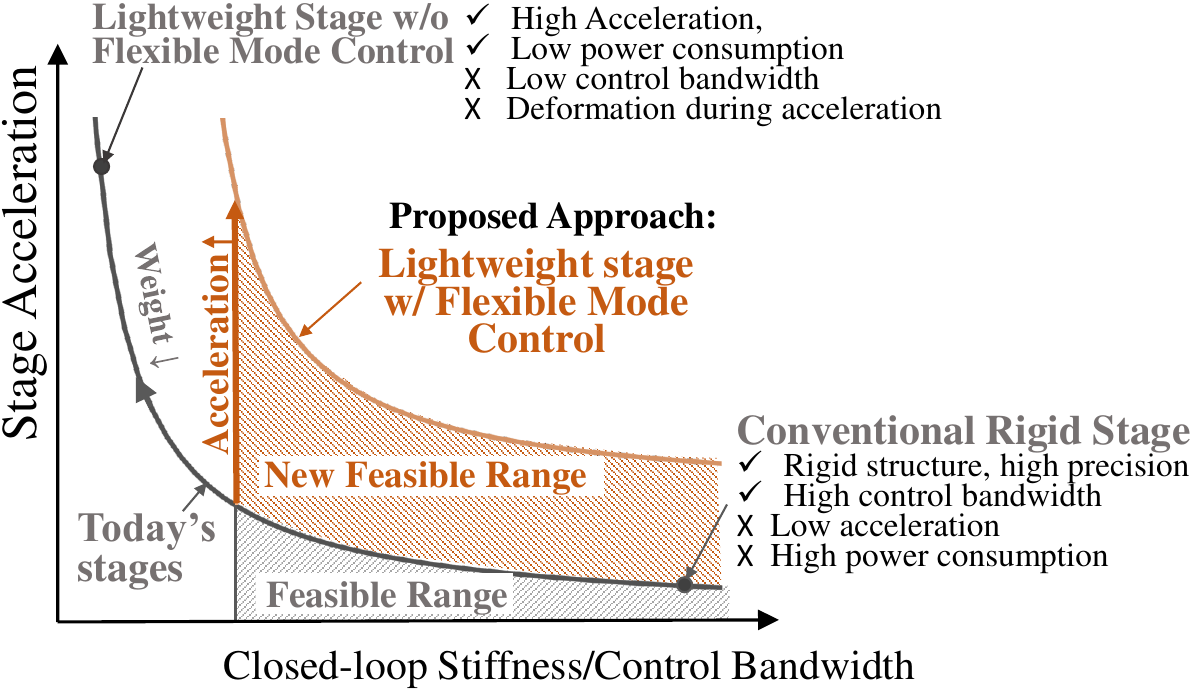}}
\vspace{-2mm}
\caption{Illustration of acceleration and bandwidth trade-off in today's precision positioning systems and motivation for the proposed lightweight stage with flexible mode control.}
\label{fig:motivation}
\end{figure}

In the past decades, a lot of research work have studied the structural design and control of lightweight precision positioning stage aiming to address the aforesaid challenge and thus improve their acceleration performance. For example, Laro~et al.~\cite{laro2010design} proposed an over-actuation method to place the actuators and sensors at the mode shapes' nodal locations to avoid exciting the flexible dynamics in the closed loop and thus increase the control bandwidth. Robbert~et al.~\cite{van2014exploiting} exploited the new design freedom from additional actuators and sensors by using a systematic robust $H_{\infty}$ control design framework to explicitly consider the flexible dynamical behavior in controller design. Despite the effectiveness, these studies only investigate the actuator/sensor placement and motion control for flexible stage but not the stage's structural design and its coupling with controller design. Recently, the combined hardware and control co-design, also referred to as control co-design (CCD), has enabled a synergistic and systematic structural-control design framework for lightweight precision motion stage to address such hardware-control design coupling. For example, Wang et al. \cite{wang2019integrated} studied the integration of structural topology optimization and actuator configuration optimization on a 3D wafer stage structure. Wu et al. \cite{JingjieCoDesign} proposed a nested CCD formulation of for lightweight motion stages with frequency specification optimized and physical limitation considered explicitly. Although significant advances have been achieved, we made a key observation that in these prior lightweight precision stages, only the stage's rigid body motion are actively controlled, and the stage's flexible dynamics are under open-loop. For such systems, the achievable control bandwidth is limited by the first structural resonance frequency. This fact enforces a \textbf{fundamental trade-off} between the stage's control bandwidth and acceleration capability, as is illustrated in Fig.~\ref{fig:motivation}. 

\begin{figure}[t!]
\centering
\subfloat{
\includegraphics[trim={0mm 0mm 0mm 0mm},clip,width =0.8\columnwidth, keepaspectratio=true]{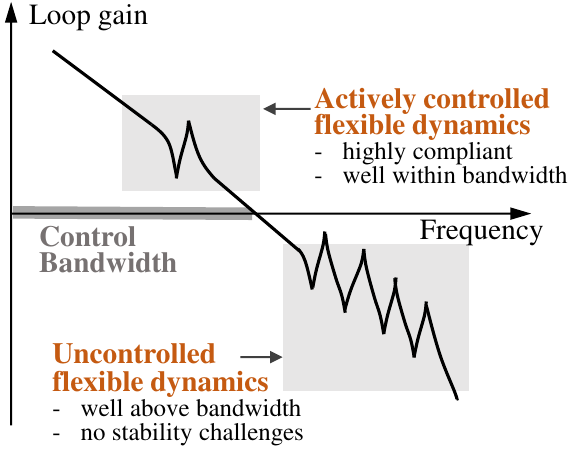}}
\vspace{-2mm}
\caption{Illustration of the  proposed lightweight stage design with active  flexible mode control.}
\vspace{-2mm}
\label{fig:proposed_cartoon}
\end{figure}

Aiming to overcome the aforementioned challenge and thus creating new lightweight precision stage with substantially improved acceleration capability without sacrificing stage control performance, this research presents a novel sequential structure and control design framework for lightweight stages with low-frequency flexible modes of the stage being actively controlled. We propose to introduce additional actuators and sensors to actively control the flexible structural dynamics of the lightweight stage to attain high control bandwidth. This concept has been explored in  \cite{van2014exploiting, proimadis2021active}, where additional actuators and sensors are used to compensate for the flexible deformations of a wafer stage. However, in these studies, the stage structures have a relatively high  stiffness, making the feedback control of flexible modes highly challenging. In addition, instability in position control can arise due to imperfect modal decoupling, which obstructs the practical application of this method.
 In order to create stage structures that can facilitate controller design, we further propose to \textbf{minimize} the resonance frequencies of the stage's mode being actively controlled  and to \textbf{maximize} the resonance frequencies of the uncontrolled modes. The target closed-loop control bandwidth lies in between the resonance frequencies as shown in Fig.~\ref{fig:proposed_cartoon}. Such a design will allow for material removal in controlled resonance mode shapes to reduce the stage's weight while enabling high control bandwidth which is limited by the uncontrolled resonance frequencies. A case study is simulated to evaluate the effectiveness of the proposed approach, where a stage weight reduction of $24\%$  is demonstrated  compared to a baseline case, which demonstrates the potential of the proposed design framework. Experimental evaluation of the designed stage's motion performance will be performed on a magnetically levitated linear motor platform for performance demonstration.

\section{Control Co-design Formulation}
\vspace{1mm}

This section presents a sequential control co-design framework to design the hardware and controller for lightweight motion stages with their low-frequency flexible modes actively controlled. First, an optimization problem that determines the stage's geometric parameters is formulated to facilitate the active control for low-frequency flexible modes. Second, another optimization that determines the actuator and sensor locations is performed to enhance the control bandwidth. Finally, feedback controllers are synthesized for the decoupled plant dynamics to control the stage's motion and low-frequency flexible dynamics. Detailed algorithms are elaborated in the following sections.

\vspace{0mm}

\subsection{A. Stage Geometry Design}
\vspace{1mm}

 To enable such a stage structure design with low-frequency flexible modes controlled, the stage's geometry optimization can be formulated as the following optimization problem: 
\vspace{-1mm}
\begin{align}  \label{eqn: shape_opt}
\begin{split}
    \min_{\theta_p}~~~&{J_m}(\theta_p),
\\
    \mathrm{s.t.}~~~&\omega_i  \leq \omega_{low}, ~~~~~~~ i=1,...,n
\\
    & \omega_j \geq \omega_{high}, ~~~~~ j=n+1,...,m 
\\
    & \theta_{p, min} \leq \theta_p \leq \theta_{p, max}.
\end{split}    
\end{align}
Here, the objective function $J_m$ represents the stage's weight, $\theta_p$ is a vector for the stage's geometric parameters,  $\omega_i$ is the $i$-th modal frequency with its corresponding vibration mode  actively controlled, and $\omega_j$ is the $j$-th resonance frequency where the corresponding mode shape is not controlled. $\omega_{low}$ is the upper bound for the actively-controlled resonance frequencies, and $\omega_{high}$ is the lower bound for the uncontrolled resonance frequencies. $\theta_{p, min}$ and $\theta_{p, max}$ are the lower and upper bounds for the stage's geometric parameter, respectively. Such an optimization process can enforce material removal in the stage's structure to allow for compliance in the actively-controlled flexible modes, and add material to stiffen the uncontrolled modes.

The selection of $\omega_{low}$ and $\omega_{high}$ are highly important and determine the designed stage's dynamic behavior.
The system's target control bandwidth must be between $\omega_{low}$ and $\omega_{high}$, and $\omega_{high}$ sets the new upper bound for the achievable control bandwidth for the lightweight precision stage with actively controlled flexible modes, as illustrated in Fig.~\ref{fig:motivation}. 

To facilitate controller design while maintaining design feasibility, the values of $\omega_{low}$ and $\omega_{high}$ need to be properly selected according to the target control bandwidth, for example $\omega_{low}\sim\frac{1}{2}\times \omega_{bw}$ and $\omega_{high}\sim5\times\omega_{bw}$, where $\omega_{bw}$ is the target bandwidth. This method, although robust, may lead to a relatively conservative stage design. To fully evaluate the feasible design range in Fig.~\ref{fig:motivation}, the value of $\omega_{high}$ needs to be swept while considering the actuator/sensor placement.

\vspace{1mm}

\subsection{B. Actuator and Sensor Placement}
\vspace{1mm}

With the stage's structure fixed, the actuator and sensor placement optimization problem for the proposed lightweight motion stage can be formulated as 
\vspace{-2mm}
\begin{align}  \label{eqn:act_opt}
\max_{\theta_a\in{D_a}}J_a(\theta_a)  = \sum_{i=1,...,n}W_{pi}(\theta_a) - \gamma\sum_{i=n+1,...,m} W_{pi}(\theta_a),
\end{align}
\vspace{-4mm}
\begin{align}  \label{eqn:sen_opt}
\max_{\theta_s\in{D_s}}J_o(\theta_s)  = \sum_{i=1,...,n}W_{oi}(\theta_s) - \gamma\sum_{i=n+1,...,m}W_{oi}(\theta_s),
\end{align}
\vspace{-6mm}

where $\theta_a$ and $\theta_s$ are vectors of actuator and sensor placement parameters, respectively; $D_a$ and $D_s$ are the design domains for actuator/sensor locations,  and $\gamma$ is a positive user-defined weighting constant. $W_{pi}$ and $W_{oi}$ are the controllability and observability grammians of $i$-th flexible mode, respectively, which can be calculated as 
\begin{align}
    W_{pi} = \frac{\| \phi_i(\theta_a)^\top B_a(\theta_a)\|_2^2}{4\zeta_i\omega_i}, \label{eq:W_p}\\
    W_{oi} = \frac{\| C_s(\theta_s)^\top \phi_i(\theta_s)\|_2^2}{4\zeta_i\omega_i}\label{eq:W_o},
\end{align}
\vspace{-6mm}

where $\phi_i$ is the mass-normalized mode shape of $i$-th flexible mode, $B_a$ and $C_s$ are the force and measurement assembling matrices, $\zeta_i$ is the modal damping ratio, and $\omega_i$ is the $i$-th resonance natural frequency. The controllability/observability grammians $W_{pi}$ and $W_{oi}$ quantitatively evaluate the controllability/observability of the corresponding flexible mode in the control system, 
which will reflect on the peak resonance magnitude in the system's frequency response.

With actuator/sensor placement optimization formulation in \eqref{eqn:act_opt} and \eqref{eqn:sen_opt}, our goal is to maximize the controllability/observability for the actively-controlled modes to reduce the required controller gain, and to minimize those of the uncontrolled modes to reduce their coupling with the control systems and thus facilitate the controller design. Notice that the value of weighting parameter $\gamma$ provides a trade-off between the two design goals: a low value in $\gamma$ emphasizes reducing the needed controller gain for actively-controlled modes, and a high value in $\gamma$ emphasizes reducing the cross-talk between uncontrolled modes and controlled modes.

\begin{figure}[t!]
\centering
\subfloat{
\includegraphics[trim={0mm 0mm 0mm 0mm},clip,width =1\columnwidth, keepaspectratio=true]{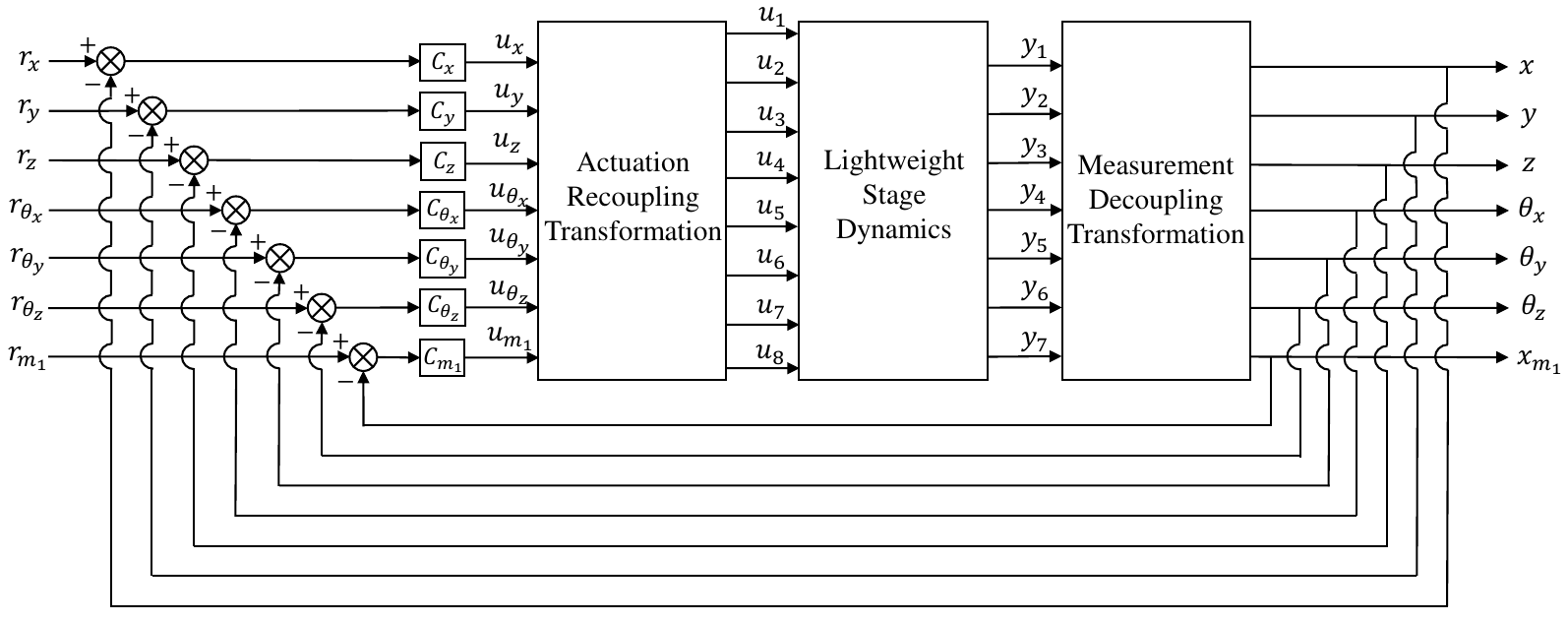}}
\vspace{-2mm}
\caption{Control block diagram for the lightweight precision positioning stage with model decoupling. }
\vspace{-5mm}
\label{fig:control_diagram}
\end{figure}

\begin{table}[t] 
\centering
\vspace{2mm}
\caption{Controller parameters  \cite{butler2011position}. }
\vspace{-6mm}
\label{table:PID_para}
    \begin{center} \begin{small}
        \begin{tabular}{ p{1.2cm} p{4cm}  p{1cm} }
        \hline
        Parameter & Description & Typical Value \\
        \hline
        $\omega_{bw}$  & Desired bandwidth [rad/s]  & --  \\
        
        
        
        $\alpha$ & PID frequency ratio  &   3 \\
        
        $K_p$  &    Proportional gain  & --   \\
        
        $\omega_i$  & Integrator frequency  & $  \omega_{bw}/ \alpha^2$  \\
        
        $\omega_d$  & Differentiator frequency  & $\omega_{bw}/\alpha$ \\
        
        $\omega_{lp}$  & Lowpass filter frequency &  $  \alpha \omega_{bw}$  \\
        
        $z_{lp}$  & Lowpass filter damping ratio  & 0.7\\
        \hline

        \end{tabular}
    \end{small} \end{center}
\vspace{-4mm}
\end{table}

\begin{figure*}[t!]
\centering
\subfloat{
\includegraphics[trim={0mm 0mm 0mm 0mm},clip,width =2\columnwidth, keepaspectratio=true]{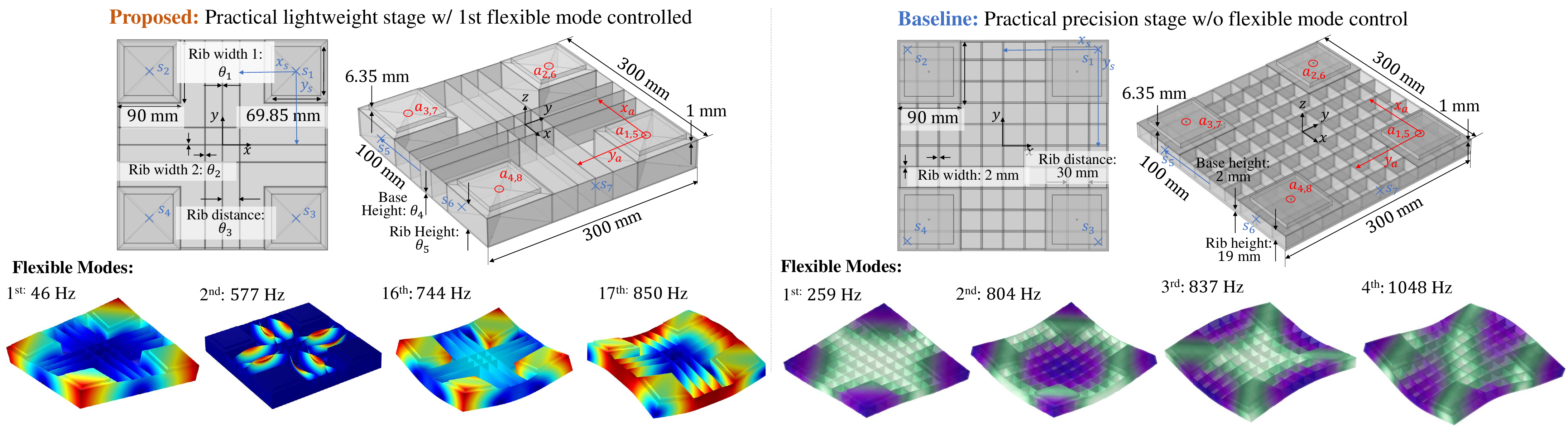}}
\vspace{-2mm}
\caption{Proposed and baseline stages. Both stages considers a permanent magnet array with $69.85~\mathrm{mm}\times69.85~\mathrm{mm}\times6.35~\mathrm{mm}$ for planar motor force generation. }
\vspace{-4mm}
\label{fig:case_2_definition}
\end{figure*}

\begin{figure*}[t!]
\centering
\subfloat{
\includegraphics[trim={0mm 0mm 0mm 0mm},clip,width =1.35\columnwidth, keepaspectratio=true]{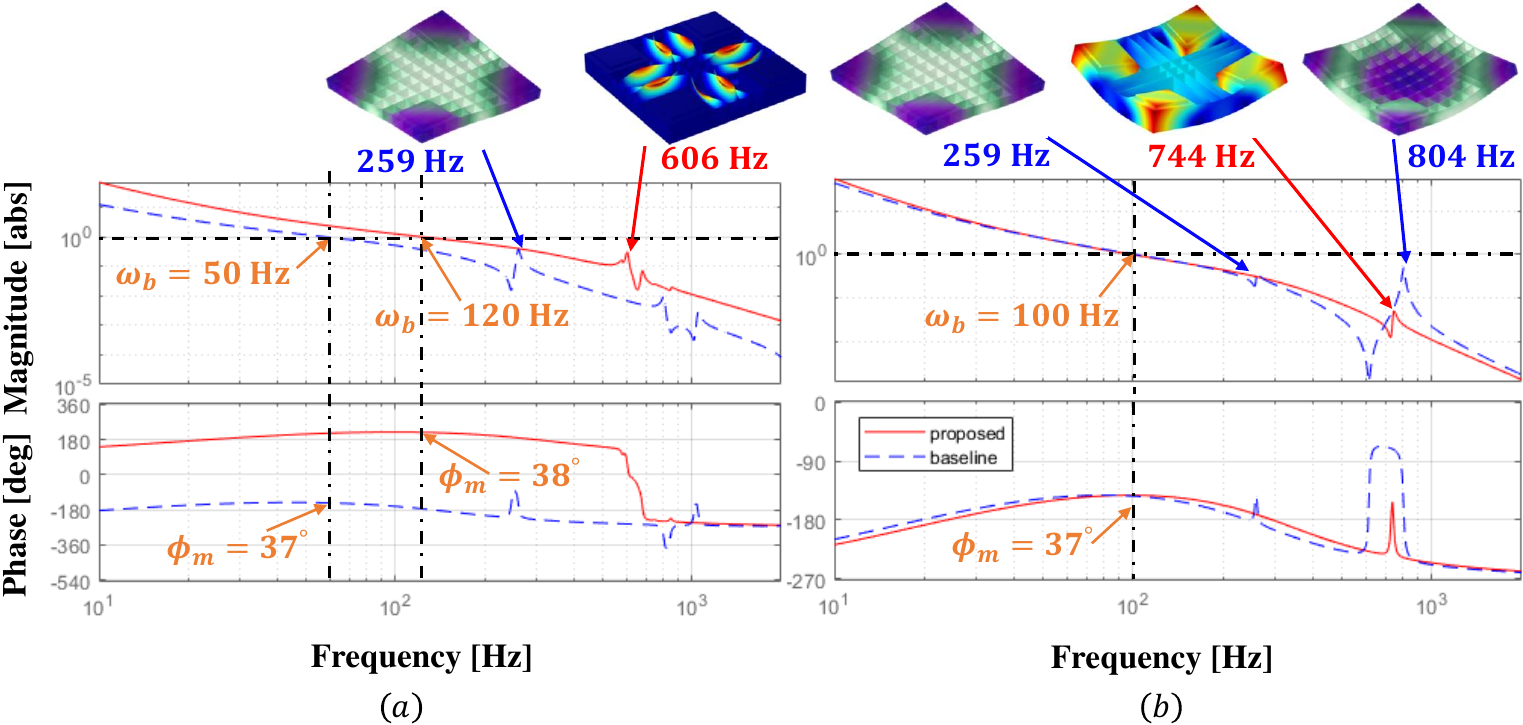}}
\vspace{-2mm}
\caption{Comparison of loop gains of the proposed design (red solid) and baseline design (blue dashed). (a) $y$-DOF (translation in the horizontal direction).  (b)  $z$-DOF (translation in the vertical direction). }

\vspace{-5mm}
\label{fig:case_2_closed_loop}
\end{figure*}

\subsection{C. Feedback Controller Design}
\vspace{1mm}

With the stage's hardware design (including both stage's structure and actuator/sensor placement) fixed, the plant dynamic can be found. Finally,  feedback controllers are designed for each motion degree of freedom to attain the target control performance.
Figure~\ref{fig:control_diagram} shows a block diagram for the control loop for a lightweight stage with all six rigid-body DOFs and one flexible mode under active control. 
Here, the lightweight stage plant dynamics $P:u\to y$
can be obtained from solving  \eqref{eqn: shape_opt}, \eqref{eqn:act_opt}, and \eqref{eqn:sen_opt}. The sensor measurements $y$ are transformed to individual DOFs via a measurement decoupling transformation. Seven single-input, single-output (SISO) feedback controllers can then be designed for seven decoupled channels 
assuming the cross-coupling between different DOFs is negligible near the target control bandwidth. For each DOF, a fixed-structure SISO controller is selected following reference \cite{franklin2002feedback} as
\begin{align}
\begin{split}   \label{eqn:PID}
    C_k(s) = K_p\Big(\frac{s+\omega_i}{s}\Big)\Big(\frac{s}{\omega_d}+1\Big)\Big(\frac{\omega_{lp}^2}{s^2+2z_{lp}\omega_{lp}s+\omega_{lp}^2}\Big),
\end{split}
\end{align}

where the controller parameters are described in Table~\ref{table:PID_para}. This controller design follows reference \cite{butler2011position} where all the controller parameters except the controller gain can be determined by a target control bandwidth $\omega_{bw}$. This approach effectively simplifies the parameter tuning process.
The proportional gain $K_p$ and the target bandwidth are determined such that the control bandwidth is maximized while satisfying a robustness criteria\cite{ortega2004systematic} of
\begin{align}  \label{eqn:robustness}
 \| S_k(s)\|_{\infty} \leq 2, k = 1, ..., n, 
\end{align}
where $S_k(s)$ is the closed-loop sensitivity function of the $k$-th channel as $S_k = (I-G_k C_k)^{-1}$. 
With the control effort signals $u_k$ for each channel computed, an actuation recoupling transformation is used to map the control signals to individual actuators.


\section{Simulation Evaluation: Magnetically Levitated Planar Stage}

To evaluate the effectiveness of the proposed framework, a case study for a practical lightweight magnetically levitated planar stage with actuator's weight and location considered is simulated and compared to a baseline design where the stage's flexible modes are not actively controlled. Then, the proposed stage design is fabricated and the experimental setup for the magnetically levitated motor is built to verify the stage's actual closed-loop motion performance. The details of the simulations and hardware are illustrated in the following sections.

\subsection{A. Simulation Evaluation}
\vspace{1mm}

Fig.~\ref{fig:case_2_definition} shows the magnetically levitated motion stage considered in this work, which is a rib-reinforced structure  made of 7075-T6 aluminum alloy of
300~mm~$\times$~300~mm in size. There are four neodymium permanent magnet arrays of 69.85~mm $\times$ 69.85~mm $\times$ 6.35~mm arranged at the corners of the stage to provide both the thrust forces for planar motion and the levitation forces. Therefore, in this case, the actuator positions are fixed so we are only able to place the sensors. The first four vibration mode shapes and corresponding resonance frequencies are also shown in Fig.~\ref{fig:case_2_definition}.  Herein, the rigid-body motion of the stage in six DOFs are actively controlled. In addition, the proposed design controls the first flexible mode while the baseline has no control on flexible modes. Besides, the placement of the sensors is optimized in the proposed stage but not in the baseline case. With a target bandwidth of 50~Hz, the baseline stage's geometric parameters are designed to constrain the first resonance frequency above 250~Hz. The geometric parameters  $\theta_p \in\mathbb{R}^{5}$  and the actuator/sensor location parameters $\theta_a = [x_a,y_a]^\top$ and $\theta_s=[x_s,y_s]^\top$ are also shown in Fig.~\ref{fig:case_2_definition}.

\begin{table}[t!] 
\centering
\renewcommand{\arraystretch}{1.0}
\caption{Case study optimal parameters}
\vspace{-4mm}
\label{table:case_2_para}
    \begin{center} \begin{small}
        \begin{tabular}{  p{2.7cm} | p{1.6cm} | p{1.8cm} }
        \hline
           & Baseline Design & Proposed Design \\ \hline
           Stage weight & 2.21 kg& 1.68 kg \\
           First res. freq. & 259~Hz & 46 Hz  \\
           2nd res. freq.  & 804~Hz & 577 Hz  \\
           $z$ motion bandwidth & 50 Hz  & 120 Hz \\
           $\theta_x$ motion bandwidth  & 100 Hz  & 100 Hz \\
           Max sensitivity  & 1.94  & 1.92 \\ 
        \hline
        \end{tabular}
            \vspace{-4mm}
    \end{small} \end{center}
\end{table}

Due to the geometric complexity of the ribbed stage structure with permanent magnets, analytical models are not sufficient to capture its structural dynamics accurately. In this work, finite element (FE) simulation (with COMSOL Multiphysics) is used to simulate the stage's spatial-temporal behavior. 
In the stage geometry optimization problem \eqref{eqn: shape_opt} formulation for the proposed stage in Fig.~\ref{fig:case_2_definition}, 
to facilitate controller design with a target control bandwidth of $\sim100~\rm{Hz}$, the values of $\omega_{low}$ and $\omega_{high}$ are selected as $50~\rm{Hz}$ and $600~\rm{Hz}$, respectively. In addition, the rib width and base height are constrained to be larger than $1~\rm{mm}$ and $0.635~\rm{mm}$ respectively for the sake of manufactuability.
With the stage geometry optimization problem \eqref{eqn: shape_opt} fully formulated, the Optimization Module in COMSOL Multiphysics is selected to solve the problem, where  an iterative method for derivative-free constrained optimization COBYLA~\cite{powell1994direct} is employed. The resultant stage resonance frequencies and mode shapes are illustrated in Fig.~\ref{fig:case_2_definition}.  Then, the sensor positioning optimization problem \eqref{eqn:sen_opt} is solved with $\gamma = 50$.

Fig.~\ref{fig:case_2_closed_loop} illustrates the loop gains of both proposed and baseline designs in $y$- and  $z$-DOFs, the performance of which are most critical. In Fig.~\ref{fig:case_2_closed_loop}a, it can be observed that with sufficient stability margins in both cases, the control bandwidth in proposed design is significantly larger than baseline design in y-DOF since the 259~Hz resonance is limiting the bandwidth. In Fig.~\ref{fig:case_2_closed_loop}b, both cases can achieve 100~Hz bandwidth. However, it can be observed that the 259~Hz resonance is still coupled in this channel but with a small peak magnitude and imperfect actuator/sensor placement or decoupling can increase the peak and thus cause stability issue. Besides, the lightly-damped 259~Hz resonance mode can be easily excited by external disturbance and thus  impair the stage's positioning precision. The weight of the proposed stage design (1.68~kg) is reduced by 24\% compared to baseline design (2.21~kg). The comparisons indicate the potential of the proposed framework to improve  stage's acceleration while maintaining high  bandwidth and  accuracy.

\begin{figure}[t!]
\centering
\subfloat{
\includegraphics[trim={0mm 0mm 0mm 0mm},clip,width =1\columnwidth, keepaspectratio=true]{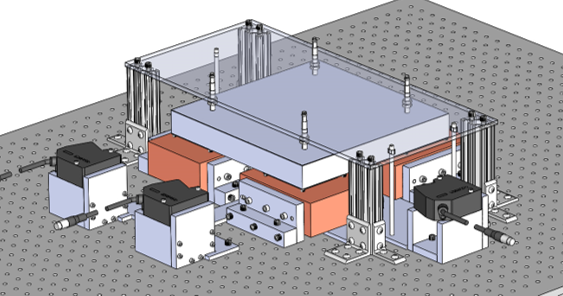}}
\vspace{-2mm}
\caption{Planar motor hardware CAD model for experimental evaluation. }
\vspace{-2mm}
\label{fig:motor_assembly}
\end{figure}

\subsection{B. Experimental Evaluation}
\vspace{1mm}

\begin{figure}[t!]
\centering
\subfloat{
\includegraphics[trim={0mm 0mm 0mm 0mm},clip,width =0.7\columnwidth, keepaspectratio=true]{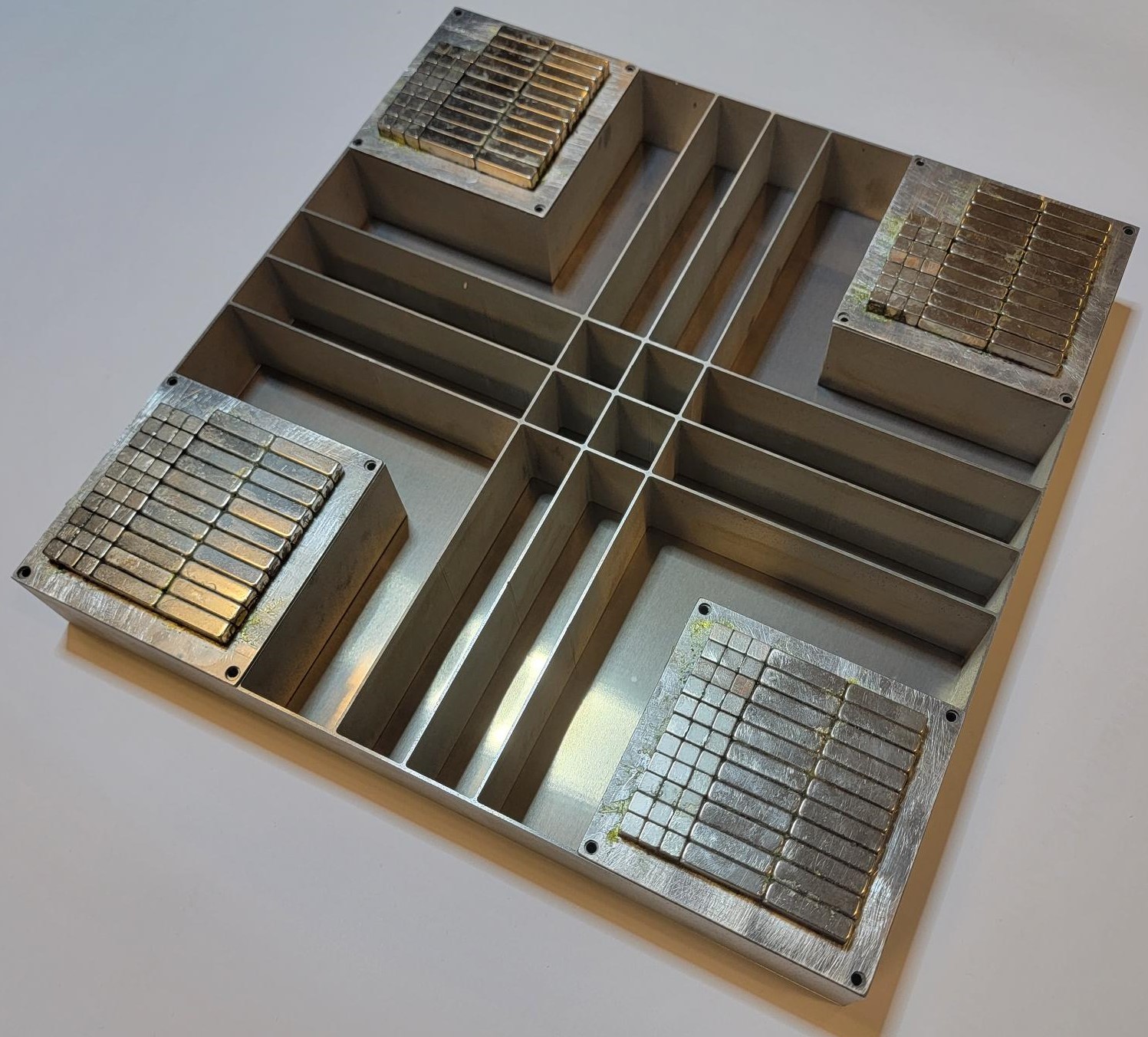}}
\vspace{-2mm}
\caption{Fabricated proposed stage hardware. }
\vspace{-2mm}
\label{fig:stage_hardware}
\end{figure}

To verify the improvement and effectiveness of our proposed framework, the designed stage of the case study is fabricated as shown in Fig.~\ref{fig:stage_hardware}. The ribbed aluminum structure and base sheet are both done by precise waterjet cutting with taper compensation. The magnet arrays at corners to provide thrust and levitation force has the pattern of traditional 4-segment-per-spatial-period Halbach array, which can generate higher magnetic field. They are made of N52 NdFeB permanent magnets with spatial period to be 25.4 mm and assembled by Loctite structural adhesive. 

\begin{figure}[t!]
\centering
\subfloat{
\includegraphics[trim={0mm 0mm 0mm 0mm},clip,width =0.8\columnwidth, keepaspectratio=true]{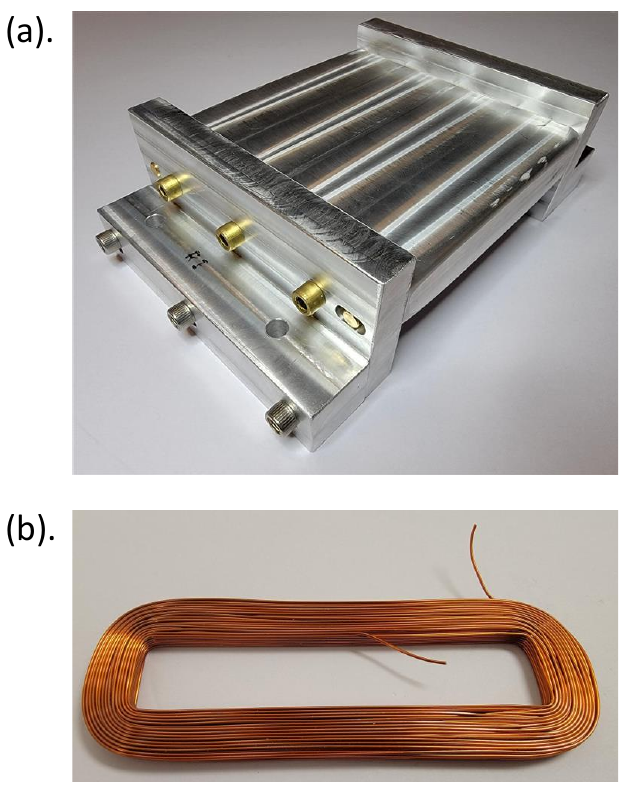}}
\vspace{-2mm}
\caption{(a) Planar motor stator. (b) Coil winding. }
\vspace{-2mm}
\label{fig:stator_coil}
\end{figure}

The overall planar motor design for actuation and sensing of the stage is shown in Fig.~\ref{fig:motor_assembly}. Four DW-AS-509-M8-390 inductive sensors are placed at the location computed from \eqref{eqn:sen_opt} to measure the vertical displacement. Three Keyence LK-H152 laser displacement sensors are used to measure the full x and y axis motions. The planar motor stators are made of Aluminum 6061 as shwon in Fig.~\ref{fig:stator_coil} (a). Their thickness are chosen to be 25.4 mm so that the magnet arrays are sufficiently far from the bottom half of the coil and the optical table while the winding resistance is not too large. To maximize the instantaneous force capability and therefore the acceleration capability, the thickness of the winding coils is chosen to be 15 mm according to parameter sweeping in the Ansys Maxwell electromagnetic simulations. The APEX PA12 power op-amp we are using has a typical voltage swing of $\pm$ 34 V and peak current of 10 A. Considering its capability, we choose AWG\#19 for the coil winding with sufficiently small total winding resistance to generate the desired peak current. One single coil winding consists of 100 turns with MWS AWG\#19 resistance-bondable magnet wire as shown in the Fig.~\ref{fig:stator_coil} (b).  

The whole hardware setup is still under construction and should be finished soon. Then, we will be able to perform the experimental evaluation on the stage's acceleration capability and closed-loop performance.

\section{Conclusions and Future work}

In this work, we have proposed and evaluated a sequential hardware-control co-design framework for lightweight precision positioning stages with high acceleration capability and high control bandwidth simultaneously. The performance and effectiveness of the framework is first evaluated by numerical simulations using a magnetically levitated planar motion  stage as case study. The significant weight reduction and improvement in control bandwidth of proposed design compared to a baseline case demonstrate the huge potential. A planar motor prototype with active flexible dynamics control is designed and now under construction in our lab. We aim to demonstrate the experimental evaluations in the near future.

\balance
\bibliography{Co_design_ASPE}

\begin{thebibliography}{23}

\bibitem{oomen2013connecting}
Oomen T, van Herpen R, Quist S, van~de Wal M, Bosgra O, Steinbuch M.
\newblock Connecting system identification and robust control for next-generation motion control of a wafer stage.
\newblock IEEE Transactions on Control Systems Technology. 2013;22(1):102--118.

\bibitem{laro2010design}
Laro DA, Boshuisen R, van Eijk J.
\newblock Design and control of over-actuated lightweight 450 mm wafer chuck.
\newblock In: 2010 ASPE Spring Topical meeting, Cambridge, Massachusetts, USA. ASPE; 2010. p. 141--144.

\bibitem{van2014exploiting}
van Herpen R, Oomen T, Kikken E, van~de Wal M, Aangenent W, Steinbuch M.
\newblock Exploiting additional actuators and sensors for nano-positioning robust motion control.
\newblock Mechatronics. 2014;24(6):619--631.

\bibitem{wang2019integrated}
Wang J, Zhang M, Zhu Y, Yang K, Li X, Wang L, et~al.
\newblock Integrated optimization of 3D structural topology and actuator configuration for vibration control in ultra-precision motion systems.
\newblock Structural and Multidisciplinary Optimization. 2019;60(3):909--925.

\bibitem{JingjieCoDesign}
Wu J, Zhou L.
\newblock Control Co-design of Actively Controlled Lightweight Structuresfor High-acceleration Precision Motion Systems.
\newblock In: American Control Conf., to appear; 2022. .

\bibitem{proimadis2021active}
Proimadis I, Custers CH, T{\'o}th R, Jansen J, Butler H, Lomonova E, et~al.
\newblock Active Deformation Control for a Magnetically Levitated Planar Motor Mover.
\newblock IEEE Transactions on Industry Applications. 2021;58(1):242--249.

\bibitem{butler2011position}
Butler H.
\newblock Position control in lithographic equipment [applications of control].
\newblock IEEE Control Sys Mag. 2011;31(5):28--47.

\bibitem{franklin2002feedback}
Franklin GF, Powell JD, Emami-Naeini A, Powell JD.
\newblock Feedback control of dynamic systems. vol.~4.
\newblock Prentice hall Upper Saddle River; 2002.

\bibitem{ortega2004systematic}
Ortega M, Rubio F.
\newblock Systematic design of weighting matrices for the H mixed sensitivity problem.
\newblock Journal of Process Control. 2004;14(1):89--98.

\bibitem{powell1994direct}
Powell MJ.
\newblock A direct search optimization method that models the objective and constraint functions by linear interpolation.
\newblock In: Adv. in opt. and num. analysis. Springer; 1994. p. 51--67.

\end{thebibliography}

\end{document}